\documentclass[english,aps,prl,showpacs,twocolumn,superscriptaddress,preprintnumbers,nofootinbib]{revtex4-1}
\global\long\def\order#1{\mathcal{O}\left(#1\right)}

\def\order#1{{\cal O}\left(#1\right)}
\def\az{\alpha_Z}

\def\gt#1{g^\text{#1}}
\def\bB{\bm{B}}
\usepackage{amsmath}
\usepackage{amssymb}
\usepackage{slashed}
\usepackage{graphicx}
\usepackage{bm}

\usepackage{epstopdf}
\begin{document}
\title{Two-loop binding corrections to the electron gyromagnetic factor}
\author{Andrzej Czarnecki}
\author{Matthew Dowling}
\affiliation{Department of Physics, University of Alberta, Edmonton, Alberta, Canada T6G 2E1}

\author{Jan Piclum}
\affiliation{Department of Physics, University of Alberta, Edmonton, Alberta, Canada T6G 2E1}
\affiliation{Theoretische Physik 1, Naturwissenschaftlich-Technische Fakult\"at, Universit\"at Siegen, 57068 Siegen, Germany}

\author{Robert Szafron}
\affiliation{Department of Physics, University of Alberta, Edmonton, Alberta, Canada T6G 2E1}
\affiliation{Physik Department T31, Technische Universit\"at M\"unchen, James Franck Stra{\ss}e~1, 85748 Garching, Germany}

\begin{abstract}
  We compute corrections to the gyromagnetic factor of an electron
  bound in a hydrogen-like ion at order $\alpha^2(Z\alpha)^5$. This
  result removes a major uncertainty in predictions for silicon and
  carbon ions, used to determine the atomic mass of the electron.
\end{abstract}
\preprint{Alberta Thy 11-17, SI-HEP-2017-18, TUM-HEP-1100/17 }
\maketitle

The Dirac equation predicts the gyromagnetic factor of a point-like
electron to be $g=2$. However, even in vacuum, self-interaction of the
electron modifies its $g$. The deviation from Dirac's prediction, a
dimensionless number known as $g-2$, can be computed in quantum
electrodynamics (QED) and expressed as a perturbation series in the
fine-structure constant $\alpha = 1/137.035\,999\,139(31)$
\cite{Mohr:2015ccw}.  This research program earned the 1965
Nobel Prize \cite{Schwinger:1948iu} and has led to the
stunning five-loop prediction of order
$\left(\frac{\alpha}{\pi}\right)^5$ \cite{Aoyama:2012wj}. In another
heroic development, the four-loop term has been calculated with very
high precision \cite{Laporta:2017okg}.  On the experimental side,
precision is so high that the electron $g-2$ is currently the best
source of $\alpha$ \cite{Gabrielse:1900zza}.

When the electron is bound in an atom, the presence of the
electrically-charged nucleus also influences the $g$-factor.  This
matters greatly for studies of trapped ions.  The $g$-factor has been
precisely measured in a range of hydrogen-, lithium- and boron-like
ions \cite{Haffner:2000zzi,Beier:2002ud,Kohler:2016yjh,%
  Sturm:2013hwa,sturm2017high,Wagner:2013iaa,Vogel:2014xha,boronlike2007}.
This experimental enterprise holds great potential for the determination
of fundamental constants and tests of the Standard Model because of
the variety of systems that can be measured \cite{Safronova:2017xyt}.
Access to nuclei with diverse values of the atomic number $Z$, and multiple electronic configurations
for a given $Z$, helps eliminate uncertainties.  Already now the
hydrogen-like carbon (combined with silicon) provides the most precise
value of the electron mass \cite{Sturm:2014bla, Zatorski:2017vro}. In
the future, also the fine-structure constant might be determined
independently from the theory and measurements of the free-electron
$g-2$ \cite{shabaev2006g,yerokhin2016g,Yerokhin:2016gxj}.

The leading binding effect on $g$ in a hydrogen-like ion, related to
the electron's motion, is known to all orders in Coulomb interactions
between the electron and the nucleus \cite{breit1928magnetic}; with
$\az=Z\alpha$,
\begin{equation}
 g^{(0)} = 2 - \frac{4}{3}\left( 1 - \sqrt{1-\az^2}  \right).
\label{eq:Breit}
\end{equation}
In our notation
$g^{(a,b)}$, $a$ denotes the power of $\frac{\alpha}{\pi}$ and $b$, if
present, the power of $\az$.  Considered together, self-interaction
and binding effects are described by a double expansion in
$\frac{\alpha}{\pi}$ and $\az$.

At one-loop level of self-interaction, $g$ is known
analytically including the very recently computed terms of order
$\frac{\alpha}{\pi}\az^5$ \cite{Pachucki:2017xfd}. Higher order terms
in $\az$ have been computed numerically
\cite{Yerokhin:2017sfg}. Numerical methods work especially well for
highly-charged ions but struggle for low values of $Z$. On the other
hand, the perturbative series in $\az$ behaves best at small $Z$. Thus
analytical and numerical methods are complementary.

\begin{figure}
\vspace*{5mm}
  \centering 
 \begin{tabular}{c@{\hspace*{4mm}}c}
   \includegraphics[width=30mm,trim=0 -.5mm 0 0cm]{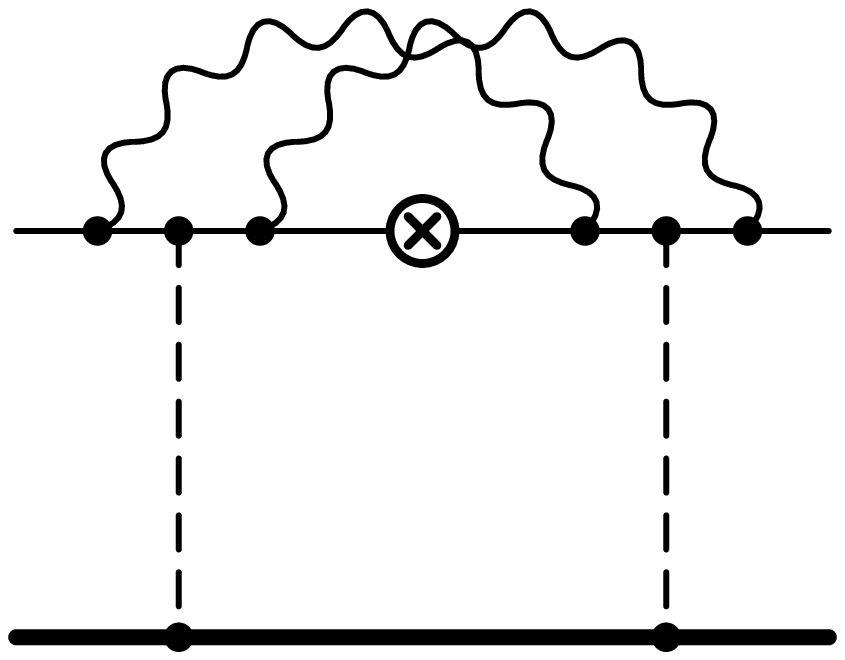} &
   \includegraphics[width=30mm]{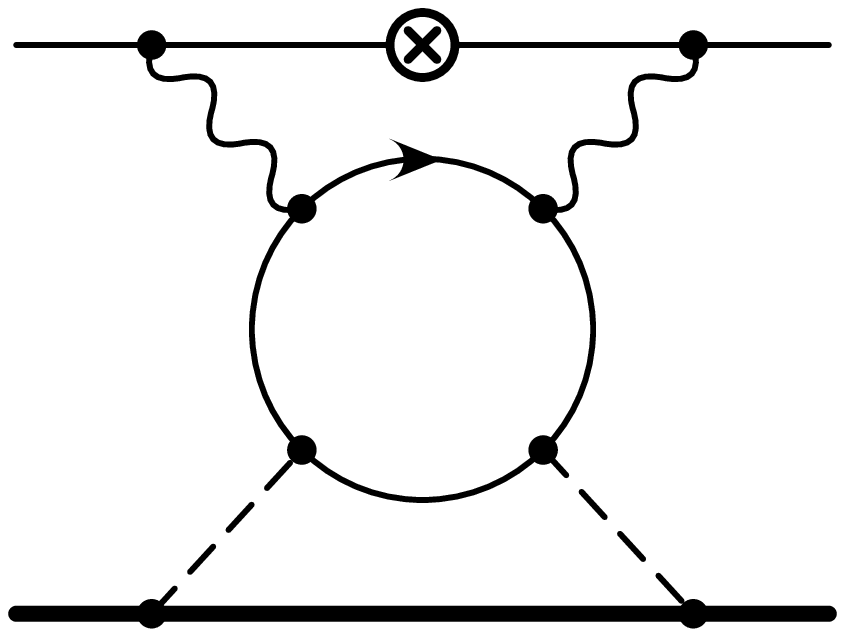} \\
  (SE) & (LBL) \\[2mm]
   \includegraphics[width=30mm]{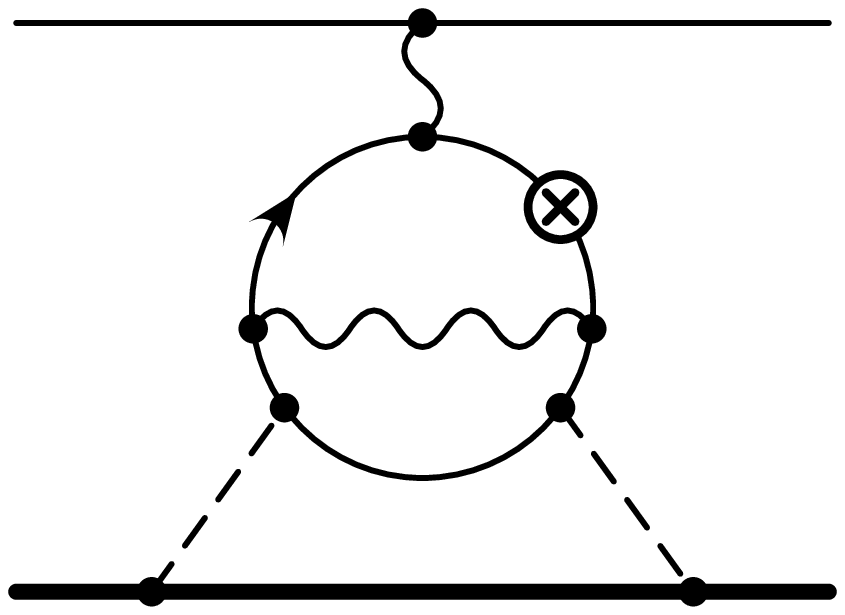} &
   \includegraphics[width=30mm]{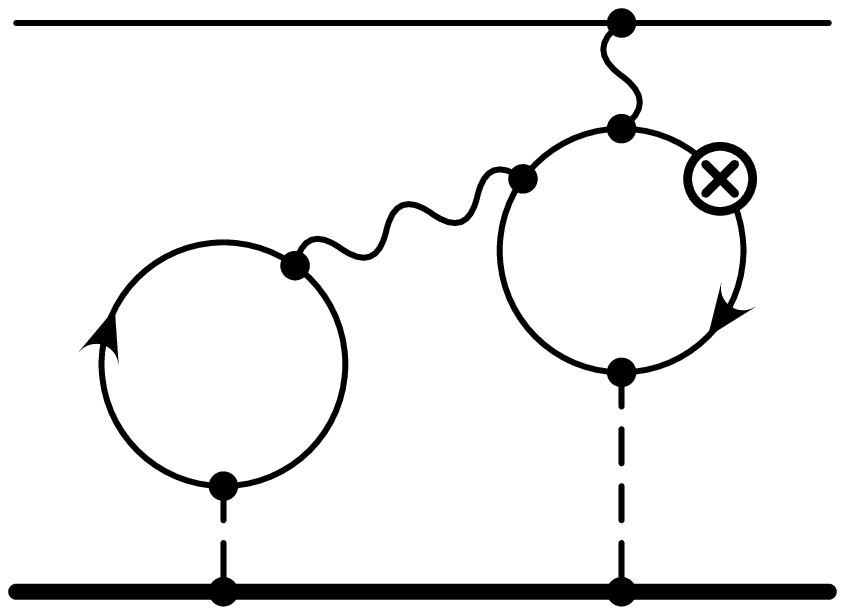} \\
   (MLPH) & (MLVP)\\
  \end{tabular}
  \caption{Examples of new contributions to the $g$-factor at order 
    $\alpha^2\az^5$. Coupling of the external magnetic field is 
    denoted with a circled cross. (SE) one of about 100 two-loop 
    self-energy diagrams; 
    (LBL) light-by-light scattering; (MLPH), (MLVP) corrections to the so-called 
    magnetic loop with a virtual photon and a vacuum polarization.}
  \label{fLoops}
\end{figure}

There is a slight tension between the two approaches. Corrections of
$\order{\frac{\alpha}{\pi}\az^5}$ arise from diagrams similar to
Fig.~\ref{fLoops}(SE), with a single self-energy (SE) loop. For an $n$S
state ($n$ is the principal quantum 
number) they are
\begin{equation}
  \label{eq:1}
  g^{(1,5)} 
= \frac{\alpha\az^{5}}{n^{3}}\left(\frac{89}{16}+\frac{8\text{ln2}}{3}\right)
=23.28 \frac{\alpha\az^{5}}{\pi n^{3}}.
\end{equation}
This result, first obtained in \cite{Pachucki:2017xfd}, is hereby
confirmed. It agrees with the coefficient $23.6(5)$ found numerically
\cite{Yerokhin:2017sfg} for the ground state $n=1$. However,
\cite{Yerokhin:2017sfg} finds a slight deviation from the scaling
$1/n^3$, contrary to \cite{Pachucki:2017xfd} and us. This seems to
indicate a poorer convergence of the numerical calculation than
assessed in \cite{Yerokhin:2017sfg}.

Two-loop self-interactions are significantly more
challenging. Numerically, they are known for the Lamb shift
\cite{Yerokhin:2003pq,Yerokhin:2004wg}, but for the $g$-factor only
some classes of diagrams have been evaluated \cite{Yerokhin:2013qma}.
Analytical results  are known  for the full set of two-loop terms up to
$\left(\frac{\alpha}{\pi}\right)^2\az^4$
\cite{Pachucki:2005px,Czarnecki:2016lzl} and for some
vacuum-polarization diagrams \cite{Jentschura:2009at}.

Effects of $\order{\left(\frac{\alpha}{\pi}\right)^2\az^5}$ are
becoming important for the determination of $m_e$. Since $g$ enters
linearly in the master formula for $m_e$ \cite{Sturm:2014bla,
  Zatorski:2017vro}, the relative error in $g$ enters directly into
the uncertainty of $m_e$. Numerically,
$\left(\frac{\alpha}{\pi}\right)^2\az^5/(g=2)$ is $4\cdot 10^{-13}$
for carbon and $3\cdot 10^{-11}$ for silicon. The current relative error
in $m_e$ is $3\cdot 10^{-11}$. Further improvements of $m_e$ hinge on
the knowledge of the coefficient of
$\left(\frac{\alpha}{\pi}\right)^2\az^5$.

In the latest determination of $m_e$ \cite{Sturm:2014bla,
  Zatorski:2017vro}, that coefficient and $m_e$ were treated as two
unknowns and were fitted to the two available results for $g$, in
H-like carbon and silicon. (This illustrates opportunities offered by
systems with various values of $Z$.) To this end, two-loop corrections were
assumed to have the following expansion in $\az$,
 \begin{eqnarray}\label{eq:corr}
 g^{(2)} &=&  \left(\frac{\alpha}{\pi}\right)^2 \left[b_{00} \left(1 + \frac{\az^2}{6} \right) 
 +\az^4 \left( b_{40} + b_{41} L \right)
 \right.\nonumber
\\
&&\left.+\az^5 b_{50} 
 +\az^6b_{63} L^3+\ldots  \right],
\label{g2}
\end{eqnarray}
The fit performed  in \cite{Sturm:2014bla, Zatorski:2017vro}
resulted in $b_{50}(\text{fit}) = -4.0(5.1)$. 
Here we directly determine contributions to $b_{50}$ beyond the already known
vacuum-polarization effects $b_{50}^\text{VP}$
\cite{Jentschura:2009at,Yerokhin:2013qma}. We write $b_{50} =
b_{50}^\text{VP}  + \Delta b_{50}$ and set out to compute 
\begin{eqnarray}
  \label{eq:2}
  \Delta b_{50} = b_{50}^\text{SE} + b_{50}^\text{LBL} + b_{50}^\text{ML},
\end{eqnarray}
where the various terms originate with Feynman diagrams shown in Fig.~\ref{fLoops}.

Numerically the largest contribution to $b_{50}$ has been expected
from $b_{50}^\text{SE}$, generated by two-loop self-energy diagrams $\Sigma^{(2)}$. They
correspond to the 19 diagrams contributing to Lamb shift  at
$\order{\alpha^2\az^5m_e}$ \cite{Pachucki:1994ega,Eides:1995ey,Dowling:2009md} (see for
example Fig.~2 in \cite{Dowling:2009md}). Each such diagram has five
virtual electron propagators as well as external electron lines, all
of which can interact with
the magnetic field. Thus we must
compute more than 100 diagrams, one example of which is
Fig.~\ref{fLoops}(SE). 

The $\az^5$ corrections require at least two photons to be exchanged
between the electron and the nucleus. The leading contribution arises
when the momentum flowing in all three loops is of the order of
electron mass (rather than for example $\order{\az m_e}$ or
$\order{\az^2 m_e}$, characteristic for so-called soft and
ultrasoft effects). In this so-called hard region, exchanges of additional Coulomb
photons are suppressed by $\az$ so there is no need to resum them. The
contribution is local from the perspective of the long-distance atomic
scale set by the Bohr radius $\sim \frac{1}{\az m_e}$ and can
be modeled by a Dirac delta potential. Corrections to this
approximation are expected to be powers and logarithms of the ratio of the
Compton wavelength of the electron and the Bohr radius, $\order{\az}$.

The contribution of such a short-distance potential to the $g$-factor
equals $\frac{4}{m_e}$ times its expectation value (its contribution
to the Lamb shift) \cite{Karsh00}; the same result follows from virial
relations \cite{shabaev2003virial}. Using the Lamb shift result
\cite{Dowling:2009md}, we find (for the ground state 1S; all our results
can be generalized to $n$S states by dividing by $n^3$)
\begin{equation}
  \label{eq:7}
  \gt{SE;Lamb} = -9.83426(5)\alpha^2 \az^5.
\end{equation}
The uncertainty in this result comes from numerical errors in the
master integrals computed in \cite{Dowling:2009md}. They affect all
quantities we compute below using these three-loop integrals.

Self-energy diagrams influence $g$ also in two other
ways. Their value depends on the energy of the electron, and that
energy is shifted by the external magnetic field $\bB$ by $\delta E =
-g\frac{q}{2m_e}\langle \bm{s} 
\rangle \cdot \bB$. The resulting correction to $g$ is called $g_3$ in
the notation of  \cite{Pachucki:2017xfd},
\begin{eqnarray}
  \label{eq:3}
  g_3 = g \left. \frac{\partial \Sigma}{\partial E}\right|_{E=m_e}
=g_3^{(1)}+g_3^{(2)},
\end{eqnarray}
where $g_3^{(1)}$ arises from the energy derivative of the one-loop
self-energy multiplied by Schwinger's correction $\Delta g =
\frac{\alpha}{\pi}$, 
\begin{equation}
  \label{eq:4}
  g_3^{(1)} = \left( 4\ln 2 - \frac{659}{64} \right) \frac{\alpha^2}{\pi}\az^5,
\end{equation}
and $g_3^{(2)}$ arises from the two-loop self-energy times the lowest order
$g \to 2$. 

Also the  wave-function of the electron is modified by the magnetic
field. This effect, together with the coupling of $\bB$ inside
self-energy diagrams like in Fig.~\ref{fLoops}(SE) provides the last SE
correction $g_4$.

To evaluate the three-loop diagrams required for $g_3^{(2)}$ and $g_4$, we use the
approach developed for the Lamb shift~\cite{Dowling:2009md}. All
three-loop diagrams are expressed in terms of 32 master integrals with
the so-called Laporta algorithm~\cite{Laporta:1996mq,Laporta:2001dd}
implemented in the program {\tt FIRE}~\cite{Smirnov:2014hma}. Results
for the master integrals and details of their computation can be found
in~\cite{Dowling:2009md}. Separately
gauge-dependent, $g_3^{(2)}$ and $g_4$  add up to give a gauge-invariant result,
\begin{equation}
  \label{eq:5}
  g_3^{(2)} + g_4 = 12.816667(72) \alpha^2 \az^5.
\end{equation}
The total SE correction is the sum of (\ref{eq:7},\ref{eq:4},\ref{eq:5})
\begin{eqnarray}
  \label{eq:SE}
  \gt{SE} = 0.58735(9) \alpha^2 \az^5.
\end{eqnarray}
We note the remarkable cancelation of sizable partial contributions to $\gt{SE}$
in this sum.

The next correction comes from light-by-light scattering diagrams, one
example of which is Fig.~\ref{fLoops}(LBL). 
They can be viewed as a term in the external-field expansion of
self-energy diagrams with a vacuum-polarization insertion. In
principle they can be calculated numerically along the lines of
Ref.~\cite{Yerokhin:2013qma}. However,  numerical treatment of
virtual (unbound) electrons is challenging. Thus we include them here
in the same manner as the SE corrections, again considering their
Lamb-shift contribution and its energy dependence. Together with the
direct magnetic field coupling effect, we find
\begin{equation}
 \label{eq:LBL}
   \gt{LBL} = -0.1724526(1) \alpha^2 \az^5.
\end{equation}

Diagrams with the external magnetic field coupling to the virtual
electron loop are expected to be small \cite{Yerokhin:2013qma}. In
this class, we consider only the so-called magnetic loop contributions
\cite{Karshenboim:2002jc}, examples of which are shown in
Fig.~\ref{fLoops}(MLPH) and (MLVP). Their origin and evaluation
differs from the diagrams we have discussed so far. Rather than
modifying the response of the electron to a given external magnetic
field, they represent the modification of the strength of the magnetic
field caused by the electrostatic field of the nucleus.

There are three types of corrections to the leading-order magnetic
loop result,
$\Delta g^\text{ML} = \frac{7}{216}\alpha\az^5$ found
in \cite{Karshenboim:2002jc}. 
Like in \eqref{eq:4},  there is a one-loop SE correction on the
main electron line that provides
\begin{equation}
  \label{eq:8}
  \gt{MLSE} =  \frac{7}{432}  \frac{\alpha^2}{\pi}\az^5 .
\end{equation}
Diagrams with a virtual photon inside the electron loop, see  for  example
Fig.~\ref{fLoops}(MLPH), give
\begin{equation}
 \label{MLPH}
  \gt{MLPH} = \left(  - \frac{7543}{16200} - \frac{303587}{10125 \pi} +      \frac{92368}{2025 \pi} \ln2 \right) \alpha^2 \az^5 .
\end{equation}
The third magnetic-loop contribution comes from inserting a second  electron loop in one of the
Coulomb-photon propagators, as in Fig.~\ref{fLoops}(MLVP), 
\begin{equation}
 \label{MLVP}
 \gt{MLVP} = \left( \frac{628}{8505 \pi } - \frac{1}{54} \right) \alpha^2 \az^5 .
\end{equation}
The total magnetic-loop correction is the sum of (\ref{eq:8},\ref{MLPH},\ref{MLVP}),
\begin{equation}
  \label{eq:ML}
  \gt{ML}  = 0.064387\dots \alpha^2 \az^5,
\end{equation}
a small effect, as expected \cite{Yerokhin:2013qma}.  
Finally, we sum  (\ref{eq:SE},\ref{eq:LBL},\ref{eq:ML}) and get the
total new correction,
\begin{equation}
  \label{eq:10}
  \Delta g^{(2,5)} = 0.479287(90) \alpha^2 \az^5,
\end{equation}
or, equivalently, a new contribution to  $b_{50}$, defined in
\eqref{eq:2},
\begin{equation}
  \label{eq:6}
  \Delta b_{50} = 4.7304(9).
\end{equation}
The error estimate in this coefficient refers to the numerical
uncertainty in the three-loop master integrals. We estimate the
additional error
due to the
yet uncalculated VP diagrams at about 13 per cent of the value in
\eqref{eq:6}, on the basis of the part we did evaluate,
eq.~\eqref{eq:ML}.
This error is presently negligible in comparison with unknown higher-order
effects and we neglect it in the numerical analysis.

We note that the magnitude of the numerical coefficient in
\eqref{eq:10} is much smaller than in the previous order, see
\eqref{eq:1}. This smallness may have complicated the experimental fit
\cite{Sturm:2014bla, Zatorski:2017vro}.  Our final result \eqref{eq:6}
has opposite sign but similar magnitude to the fitted result
\cite{Sturm:2014bla, Zatorski:2017vro},
$b_{50}(\text{fit}) = -4.0(5.1)$.  The difference between central
values is 1.7 times the error assigned to the fit.  Of course, the fit
includes VP effects that we have not considered, but the known VP
contributions are predominantly positive \cite{Yerokhin:2013qma} and
increase our discrepancy with the fit. On the other hand, the fit was
done before  the $\order{\az^4}$ LBL effect was computed
\cite{Czarnecki:2016lzl}; including it may improve the agreement \cite{HarmanPriv}.

How does the new correction in eq.~\eqref{eq:10} influence the
determination of the electron mass? In case of the carbon ion, the
relative size of the change in $g$ is $2\cdot 10^{-12}$. The atomic
electron mass is directly proportional to $g$ so this change increases
$m_e$ by the same relative amount, well below the current error of
$3\cdot 10^{-11}$ \cite{Sturm:2014bla, Zatorski:2017vro}; in absolute
terms, by about $10^{-15}$ atomic mass units.

This stability of the electron mass should not be taken for
granted. Partial results such as \eqref{eq:5} correspond to relative
shifts as large as $5\cdot 10^{-11}$, larger than the current uncertainty in
$m_e$. They have, however, been canceled by other effects such as
$\gt{SE;Lamb}$ in \eqref{eq:7}.

In Table \ref{tab:t1} we summarize results for hydrogen-like helium,
carbon and silicon ions. The first line corresponds to Breit's formula
\eqref{eq:Breit}. The second line includes additionally all known
corrections preceding this work, as given in \cite{Zatorski:2017vro}:
finite nuclear size \cite{zatorski2012nuclear}; one-loop QED
corrections obtained by combining numerical and analytical
computations,
\cite{Grotch:1970zza,Eides:1997sq,Czarnecki:2000uu,Karshenboim:2001ej,%
Yerokhin:2002pt,Lee:2004vb,Yerokhin:2004zz,Pachucki:2004si,%
Pachucki:2005px,Karshenboim:2002jc,yerokhin2013nuclear,Yerokhin:2017sfg};
two-loop QED corrections evaluated up to order $\az^4$
\cite{Yerokhin:2004zz,Pachucki:2004si,Czarnecki:2016lzl}; higher order
terms for diagrams with vacuum polarization insertions
\cite{Yerokhin:2013qma}; three and more QED loops up to order $\az^2$
\cite{Aoyama:2012wj}; recoil and radiative recoil
\cite{Shabaev:2002gg,Beier200079,Pachucki:2008zz} (see also
\cite{Glazov:2014oha}); nuclear polarizability
\cite{nefiodov2002nuclear} and susceptibility
\cite{Jentschura:2005vq}; as well as leading weak
\cite{Czarnecki:1995sz} and hadronic effects
\cite{Nomura:2012sb,Kurz:2014wya}.

For carbon and silicon the theoretical error in the second row of
Table \ref{tab:t1} is dominated by the finite nuclear size effects,
whose error is respectively $\delta g = 7 \times 10^{-13}$ and
$\delta g=31 \times 10^{-12}$. For helium, the main uncertainty comes
from the poorly known one-loop corrections extrapolated from numerical
computation.

We estimate the remaining error due to still missing higher order
terms by the leading logarithm
$\left(\frac{\alpha}{\pi}\right)^2\az^{6}\ln^3 \az^{-2}$. In carbon
and silicon, this logarithmic term exceeds other uncertainties as well
as our correction. It also exceeds the uncertainty estimated in
\cite{Zatorski:2017vro}, where for Si ion the error was
$139 \cdot 10^{-12}$. 

It may seem ironic that after all the effort of computing $\Delta
b_{50}$ our predictions for $g$ have larger
uncertainties than before our work. We believe however that this
conservative treatment of unknown higher orders is necessary. 

In \cite{Sturm:2014bla}, $b_{50}$ was fitted and the
higher-order terms were neglected. This procedure raises some
concerns since the next term, $b_{63}$, is logarithmically
enhanced. For small $Z$, that enhancement may overshadow the $b_{50}$
term. Yet, with experimental results available only for carbon and
silicon it was impossible to constrain that higher-order term. Now
that the $b_{50}$ term is at hand, one can fit the logarithmically
enhanced term $b_{63}$ while neglecting parametrically suppressed
terms starting from $b_{62}$.

In this way, our result will help derive full benefit from future
measurements of $g$-factors, an effort we admire and encourage.

\section*{Acknowledgments}
We thank Zoltan Harman for helpful discussions.
We thank Jorge Mond\'ejar for collaboration in the early stages of
this project. The loop diagrams were calculated with {\tt
  FORM}~\cite{Vermaseren:2000nd}. 
This research was supported by Science and Engineering Research Canada
(NSERC) and by the Munich Institute for Astro- and Particle Physics
(MIAPP) of the DFG cluster of excellence ``Origin and Structure of the
Universe.''

\begin{widetext}

\begin{table}[h]
\centering
\caption{\label{tab:t1} Bound electron $g$-factor for helium, carbon
  and silicon ions. The error related to missing higher order
  contributions is estimated by
 $\left(\frac{\alpha}{\pi}\right)^2\az^{6}\ln^3 \az^{-2}$ .}

\begin{ruledtabular}
\begin{tabular}{lllll}
Contribution & $^{4}\textrm{He}^{+}$                  & $^{12}\textrm{C}^{5+}$              & $^{28}\textrm{Si}^{13+}$        & Source               \\
\hline 
Dirac/Breit value &\phantom{$-$}1.999\,857\,988 825\,37(6)    & \phantom{$-$}1.998\,721\,354\,392\,1(6)    & \phantom{$-$}1.993\,023\,571\,557(3)   & \cite{breit1928magnetic}    \\
 + other known corrections 
 
&\phantom{$-$}2.002\,177\,406\,711\,41(55) &
 \phantom{$-$}2.001\,041\,590\,168\,6(12)   & \phantom{$-$}1.995\,348\,957\,825(39) & \cite{Zatorski:2017vro,Yerokhin:2017sfg}\footnote{The only change with respect to \cite{Zatorski:2017vro} is that we used the new evaluation of \cite{Yerokhin:2017sfg} for the one-loop corrections.}      \\
								 &                   		     	   &					      &		     		   	     &   \\
 $ g^{\textrm{SE}}   $           	& \phantom{$-$}0.000\,000\,000\,000\,02   &\phantom{$-$}0.000\,000\,000\,005\,0    &\phantom{$-$}0.000\,000\,000\,348     	     & (this work) \\
 $ g^{\textrm{LBL}}  $            	&           $-$0.000\,000\,000\,000\,01   &          $-$0.000\,000\,000\,001\,5    &          $-$0.000\,000\,000\,102      & (this work) \\
 $ g^{\textrm{ML}}   $			& \phantom{$-$}0.000\,000\,000\,000\,00	&\phantom{$-$}0.000\,000\,000\,000\,6    &\phantom{$-$}0.000\,000\,000\,038      & (this work) \\
 \hline
$\left(\frac{\alpha}{\pi}\right)^2\az^{6}\ln^3 \az^{-2}$ 			& \phantom{$-$}0.000\,000\,000\,000\,00(3)  &\phantom{$-$}0.000\,000\,000\,000\,0(93)&\phantom{$-$}0.000\,000\,000\,000(583) &	\\
 \hline
 Total 					&\phantom{$-$}2.002\,177\,406\,711\,42(55)  &\phantom{$-$}2.001\,041\,590\,172\,7(94)&\phantom{$-$}1.995\,348\,958\,109(584) &
\end{tabular}
\end{ruledtabular}

\end{table}
\end{widetext}

%

\end{document}